\documentclass[a4paper, 12pt, oneside]{article}
\usepackage[cp1251]{inputenc}
\usepackage[english, russian]{babel}
\usepackage{ graphics,amsfonts, amsmath,bm,amssymb, array, eucal}
\usepackage[dvips]{graphicx}
 {

\begin{document}
\thispagestyle{empty} \large
\renewcommand{\abstractname}{}
\renewcommand{\abstractname}{Abstract }
\renewcommand{\refname}{\begin{center} REFERENCES\end{center}}
\newcommand{\mc}[1]{\mathcal{#1}}
\newcommand{\E}{\mc{E}}

 \begin{center}
\bf Transverse electric conductivity and dielectric permeability in quantum
degenerate collisional plasma with variable collision frequency
in Mermin's approach
\end{center}\medskip
\begin{center}
  \bf A. V. Latyshev\footnote{$avlatyshev@mail.ru$} and
  A. A. Yushkanov\footnote{$yushkanov@inbox.ru$}
\end{center}\medskip

\begin{center}
{\it Faculty of Physics and Mathematics,\\ Moscow State Regional
University, 105005,\\ Moscow, Radio str., 10--A}
\end{center}\medskip

\begin{abstract}
Formulas for transverse conductance and dielectric permeability in quantum
degenerate collisional plasma  with arbitrary variable collision frequency
in Mermin's approach are deduced.
Frequency of collisions of particles depends arbitrarily on a wave vector.
For this purpose the kinetic  Shr\"{o}dinger---Boltzmann equation
with collision integral of relaxation type in momentum space is applied.
The case of degenerate Fermi plasma is allocated and investigated.
The special case of frequency of collisions
proportional to the module of a wave vector is considered.
The graphic analysis of the real and imaginary parts  of
dielectric function is made.


\end{abstract}

\begin{center}
{\bf 1. Introduction}
\end{center}

In  Klimontovich and Silin's work  \cite{Klim} expression
for longitudinal and trans\-verse dielectric permeability of quantum
collisionless plasmas has been re\-cei\-ved.

Then in Lindhard's work \cite{Lin} expressions
has been received  also for the same characteristics of quantum
collisionless plasma.

By Kliewer and Fuchs \cite{Kliewer} it has been shown, that
direct generalisation of formulas of Lindhard  on a case of collisionless
plasmas, is incorrectly.
This lack for the longitudinal dielectric
permeability has been eliminated in work of Mermin \cite{Mermin} for
collisional plasmas.
In this work of Mermin \cite{Mermin} on the basis of the analysis
of a nonequilibrium matrix
density in $ \tau $-approach expression for
longitudinal dielectric permeability of quantum collisional plasmas
in case of constant frequency of collisions of particles of plasma
has been announced.

For collisional plasmas correct formulas longitudinal and transverse
electric conductivity and dielectric permeability are received
accordingly in works \cite{Long} and \cite{Trans}. In these works
kinetic  Wigner---Vlasov---Boltzmann equation
in relaxation approximation in coordinate space was used.

In work \cite{Trans2} the formula for the transverse electric
conductivity of quan\-tum collisional plasmas with use of the kinetic
Shr\"{o}dinger---Boltzmann equation in Mermin's approach  (in space of
momentum) has been deduced.

In work \cite{Long2} the formula for the longitudinal dielectric
permeability of quantum collisional plasmas with use of the kinetic
Shr\"{o}dinger---Boltzmann equation in approach of Mermin (in space of
momentum) with any variable frequency of collisions depending from
wave vector  has been deduced.

In our work \cite{Lat2007} formulas for longitudinal and transverse
electric con\-duc\-ti\-vity in the classical collisional
gaseous (maxwellian) plasma with frequency of collisions
of plasma particles proportional to the
module particles velocity  have been deduced.

Research of
skin-effect in classical collisional gas plasma with frequency
of collisions proportional to the module particles velocity
has been carried out in work \cite{Lat2006}.

In our works \cite{Long3} and \cite{Long4} dielectric permeability
in quantum collisional plasma with frequency of collisions
proportional to the module
of a wave vector has been investigated. The case of degenerate plasmas
was studied in work \cite{Long3}. The case of non-degenerate
and maxwellian plasmas has been investigated in work \cite{Long4}.

Let's notice, that interest to research of the phenomena
in quantum plasma grows in last years \cite{Manf} -- \cite{Ropke}.

In the present work formulas for transverse conductivity and dielectric
permeability in quantum
degenerate collisional plasma  with arbitrary vari\-able collision frequency
in Mermin's approach are deduced.
Frequency of collisions of particles depends arbitrarily on a wave vector.

For this purpose the kinetic  Shr\"{o}dinger---Boltzmann equation
with colli\-sion integral of relaxation type in momentum space is applied.
The case of degenerate Fermi plasma is allocated and investigated.
The special case of frequency of collisions
proportional to the module of a wave vector is considered.
The graphic analysis of the real and imaginary parts  of
dielectric function is made.

\begin{center}
  \bf 1.  Kinetic Schr\"{o}dinger---Boltzmann equation for density matrix
\end{center}

Let the vector potential of an electromagnetic field is
harmonious, i.e. changes as
$
{\bf A}={\bf A}({\bf r})\exp(-i \omega t).
$

We consider transverce conductivity. Therefore the following relation
is carried out
$
{\rm  div}{\bf A}(\mathbf{r},t)=0.
$

Communication between vector potential and intensity of electric field
is given by following expression
$$
{\bf A}({\bf q})=-\dfrac{ic}{\omega}\;{\bf E}({\bf q}).
$$

The equilibrium matrix of density has the following form
$$
{\tilde \rho}=\dfrac{1}{1+\exp\dfrac{H-\mu}{k_BT}}.
$$

Here $T $ is the temperature, $k_B$ is the Boltzmann constant,
$\mu$ is the chemical potential of plasma, $H $ is the Hamiltonian.
Further we will be consider  dimensionless chemical potential
of plasma  $ \alpha =\mu/k_BT $.

In linear approach Hamiltonian has the following form
$$
H=\dfrac{({\bf p}-({e}/{c}){\bf A})^2}{2m}=
\dfrac{{\bf p}^2}{2m}-\dfrac{e}{2mc}({\bf p}{\bf A}+{\bf A} {\bf
p}).
$$

Here $\mathbf{p}$ is the momentum operator, $\mathbf{p}=-i\hbar \nabla$,
$e$ and $ m$ are the charge and mass of electron, $c$ is the velocity of light.

Hence, we can present this Hamiltonian  in the form of the sum
two operators  $H=H_0+H_1$,
where
$$
H_0=\dfrac{{\bf p}^2}{2m},
\qquad H_1=-\dfrac{e}{2mc}({\bf p}{\bf A}+{\bf A} {\bf p}).
$$

We take the kinetic equation for the density matrix in $ \tau $--
approximation
$$
i\hbar \dfrac{\partial \rho}{\partial t}=[H,\rho]+
\dfrac{i\hbar}{\tau}({\tilde\rho}-\rho).
\eqno{(1.1)}
$$

Here $\nu=1/\tau$ is the effective collisional frequency of plasma particles,
$\tau$ is the characteristic time between two consecutive collisions,
$\hbar$ is the Planck's constant, $[H,\rho]=H\rho-\rho H$ is the
commutator, $\mathbf{\tilde{\rho}}$ is the equilibrium matrix density.

Generally frequency of collisions $\nu$ should depend from
electron mo\-men\-tum ${\bf p}$ (or a wave vector $ {\bf k} $):
$\nu =\nu ({\bf k})$.

Considering the requirement Hermitian character the equation (1.1) on
the density matrix it is necessary to rewrite in the form
$$
i\hbar \dfrac{\partial \rho}{\partial t}=[H,\rho]+i\hbar\dfrac{\nu({\bf k})}{2}
(\tilde{\rho}-\rho)+(\tilde{\rho}-\rho)i\hbar\dfrac{\nu({\bf k})}{2}.
\eqno{(1.2)}
$$

In linear approach on an external field we search for a density matrix in
the form
$$
{\rho}={\tilde \rho}_0+{\rho}_1.
\eqno{(1.3)}
$$

Here $ {\rho} _1$ is the correction (perturbation) to the equilibrium
density matrix, caused by presence of an electromagnetic field,
$ \tilde {\rho} _0$ is the equilibrium matrix of density,
corresponds to the "equilibrium" \, Hamilton operator $H_0$.

We present the equilibrium matrix density $\tilde{\rho}$ in the following
form
$$
\tilde{\rho}=\tilde{\rho}_0+\tilde{\rho}_1.
\eqno{(1.4)}
$$

We consider the commutator $[H, \tilde{\rho}]$.
In linear approximation this com\-mu\-tator is equal
$$
[H, {\tilde \rho}\,]=[H_0, {\tilde \rho}_1]+[H_1, {\tilde\rho}_0]
\eqno{(1.5)}
$$
and
$$
[H, {\tilde \rho}\,]=0.
\eqno{(1.6)}
$$

For commutators from right side of equality (1.5) we obtain
$$
\langle\mathbf{k}_1|[H_0, \tilde{\rho}_1]|\mathbf{k}_2\rangle=
\big(\E_{\mathbf{k}_1}-\E_{\mathbf{k}_2}\big)
\langle\mathbf{k}_1|\tilde{\rho}_1|\mathbf{k}_2\rangle=
\big(\E_{\mathbf{k}_1}-\E_{\mathbf{k}_2}\big)
\tilde{\rho}_1(\mathbf{k}_1-\mathbf{k}_2),
\eqno{(1.7)}
$$
and
$$
\langle\mathbf{k}_1|[H_1, \tilde{\rho}_0]|\mathbf{k}_2\rangle=
\langle\mathbf{k}_1|H_1\tilde{\rho}_0|{\bf k}_2\rangle-
\langle\mathbf{k}_1|\tilde{\rho}_0H_1|{\bf k}_2\rangle=
$$
$$
=\dfrac{e}{2mc}\big(f_{\mathbf{k}_1}-f_{\mathbf{k}_2}\big)
(\mathbf{k}_1+\mathbf{k}_2)\langle{\bf k}_1|{\bf A}|{\bf k}_2\rangle=$$$$
=\dfrac{e}{2mc}\big(f_{\mathbf{k}_1}-f_{\mathbf{k}_2}\big)
(\mathbf{k}_1+\mathbf{k}_2)\mathbf{A}(\mathbf{k}_1-\mathbf{k}_2),
\eqno{(1.8)}
$$
where
$$
f_\mathbf{k}=\dfrac{1}{1+\exp (\beta \E_{\mathbf{k}}-\alpha)},\qquad
\E_{\mathbf{k}}=\dfrac{\hbar^2\mathbf{k}^2}{2m}, \qquad
\mathbf{p}=\hbar \mathbf{k}.
$$

From relations (1.4)--(1.8) follows that
$$
\tilde{\rho}_1(\mathbf{k}_1-\mathbf{k}_2)=
-\dfrac{e\hbar}{2mc}\dfrac{f_{\mathbf{k}_1}-
f_{\mathbf{k}_2}}{\E_{\mathbf{k}_1}-\E_{\mathbf{k}_2}}
(\mathbf{k}_1+\mathbf{k}_2)\mathbf{A}(\mathbf{k}_1-\mathbf{k}_2).
\eqno{(1.9)}
$$

By means of equalities (1.3)--(1.5) we linearize the kinetic equation (1.2).
We receive the following equation
$$
i\hbar \dfrac{\partial \rho_1}{\partial t}=[H_0,\rho_1]+$$$$+
[H_1, \tilde{\rho}_0]+i\hbar\dfrac{\nu({\bf k}_1)}{2}(\tilde{\rho}_1-\rho_1)
+(\tilde{\rho}_1-\rho_1)i\hbar\dfrac{\nu({\bf k}_2)}{2}.
\eqno{(1.10)}
$$

We notice that the perturbation $\rho_1\sim \exp(-i\omega t)$,
then from equation (1.10) we receive
$$
\hbar \omega \langle{\bf k}_1|\rho_1|{\bf k}_2\rangle=
\langle{\bf k}_1|[H_0,\rho_1]|{\bf k}_2\rangle+
\langle{\bf k}_1|[H_1,\rho_0]|{\bf k}_2\rangle+$$$$+
i\hbar \dfrac{\nu({\bf k}_1)}{2}
\langle{\bf k}_1|\bar \rho_1-\rho_1|{\bf k}_2\rangle+
\langle{\bf k}_1|\bar \rho_1-\rho_1|{\bf k}_2\rangle
i\hbar \dfrac{\nu({\bf k}_2)}{2}.
$$

We will enter the designation
$$
\bar\nu({\bf k}_1,{\bf k}_2)=\dfrac{\nu({\bf k}_1)+\nu({\bf k}_2)}{2}
$$
and rewrite the
previous equation in the form
$$
\hbar [\omega +i\hbar\bar\nu({\bf k}_1,{\bf k}_2)]
\langle{\bf k}_1|\rho_1|{\bf k}_2\rangle=
\langle{\bf k}_1|[H_0,\rho_1]|{\bf k}_2\rangle+
\langle{\bf k}_1|[H_1,\tilde\rho_0]|{\bf k}_2\rangle+
$$
$$
+i\hbar \bar\nu({\bf k}_1,{\bf k}_2)
\langle{\bf k}_1|\tilde{\rho}_1|{\bf k}_2\rangle.
$$

Using equalities (1.7) -- (1.9), we will transform this equation to
the following form
$$
\{\E_{{\bf k}_1}-\E_{{\bf k}_2}-
\hbar [\omega +i\hbar\bar\nu({\bf k}_1,{\bf k}_2)]\}
\langle{\bf k}_1|\rho_1|{\bf k}_2\rangle=
$$\medskip
$$
=-\dfrac{e\hbar}{2mc}(f_{{\bf k}_1}-f_{{\bf k}_2})
\dfrac{\E_{{\bf k}_1}-\E_{{\bf k}_2}-i\hbar\bar\nu({\bf k}_1,{\bf k}_2)}
{\E_{{\bf k}_1}-\E_{{\bf k}_2}}({\bf k}_1+{\bf k}_2)
\langle{\bf k}_1|{\bf A}|{\bf k}_2\rangle.
$$\medskip

Now from the previous equation we find
$$
\langle{\bf k}_1|\rho_1|{\bf k}_2\rangle=-\dfrac{e\hbar}{2mc}\Xi({\bf k}_1,
{\bf k}_2)(f_{{\bf k}_1}-f_{{\bf k}_2})({\bf k}_1+{\bf k}_2)
\langle{\bf k}_1|{\bf A}|{\bf k}_2\rangle.
\eqno{(1.11)}
$$

Here
$$
\Xi({\bf k}_1,{\bf k}_2)=\dfrac{\E_{{\bf k}_1}-\E_{{\bf k}_2}-
i\hbar\bar\nu({\bf k}_1,{\bf k}_2)}
{(\E_{{\bf k}_1}-\E_{{\bf k}_2})\{\E_{{\bf k}_1}-\E_{{\bf k}_2}-
\hbar [\omega +i\hbar\bar\nu({\bf k}_1,{\bf k}_2)]\}}.
$$ \medskip

In equation (1.11) we will put $\mathbf{k}_1=\mathbf{k}$,
$\mathbf{k}_2=\mathbf{k}-\mathbf{q}$.

Then
$$
\langle\mathbf{k}_1|\rho_1|\mathbf{k}_2\rangle=
\langle\mathbf{k}|\rho_1|\mathbf{k}-
\mathbf{q}\rangle=\rho_1({\bf q})=
$$ \hspace{0.2cm}
$$
=-\dfrac{e\hbar}{mc}
\Xi({\bf k},{\bf k-q})(f_{{\bf k}}-f_{{\bf k-q}})
\mathbf{k}\mathbf{A}(\mathbf{q}).
\eqno{(1.12)}
$$\hspace{0.3cm}

Here
$$
\Xi({\bf k},{\bf k-q})=\dfrac{\E_{{\bf k}}-\E_{{\bf k-q}}-
i\hbar\bar\nu({\bf k},{\bf k-q})}
{(\E_{{\bf k}}-\E_{{\bf k-q}})\{\E_{{\bf k}}-\E_{{\bf k-q}}-
\hbar [\omega +i\hbar\bar\nu({\bf k},{\bf k-q})]\}}.
$$ \medskip

\begin{center}
  \bf 2. Current density
\end{center}

The current density ${\bf j}({\bf q})$  is defined as
$$
{\bf j}({\bf q},\omega)=e\int \dfrac{d{\bf k}}{8\pi^3m}\left\langle{\bf k}+
\frac{\mathbf{q}}{2}\left|({\bf p}-\frac{e}{c}{\bf A})
\rho+\rho ({\bf p}-\frac{e}{c}{\bf A} \big)\right|{\bf k}-
\frac{\mathbf{q}}{2}\right\rangle.
\eqno{(2.1)}
$$ \medskip

After substitution (1.3) in integral from (2.1), we have
$$
\left\langle{\bf k}+
\frac{\mathbf{q}}{2}\left|({\bf p}-\frac{e}{c}{\bf A})
\rho+\rho ({\bf p}-\frac{e}{c}{\bf A} \big)\right|{\bf k}-
\frac{\mathbf{q}}{2}\right\rangle=
$$
$$
=
\left\langle{\bf k}+
\frac{\mathbf{q}}{2}\left|{\bf p}\rho_1+\rho_1\mathbf{p}-\frac{e}{c}
({\bf A}\tilde{\rho}_0+\tilde{\rho}_0\mathbf{A})\right|{\bf k}-
\frac{\mathbf{q}}{2}\right\rangle.
$$ \medskip

It is easy to show, that
$$
\left\langle{\bf k}+\dfrac{\mathbf{q}}{2}\Big|{\bf p}\rho_1+
\rho_1\mathbf{p}\Big|{\bf k}-
\dfrac{\mathbf{q}}{2}\right\rangle
=2\hbar\mathbf{k}\tilde\rho_0(\mathbf{q}),
$$

$$
\left\langle{\bf k}+
\dfrac{\mathbf{q}}{2}\Big|{\bf A}\tilde{\rho}_0+
\tilde{\rho}_0\mathbf{A}\Big|{\bf k}-\dfrac{\mathbf{q}}{2}\right\rangle=
\mathbf{A}(\mathbf{q})\Big[\tilde{\rho}_0\Big(\mathbf{k}+
\dfrac{\mathbf{q}}{2}\Big)+\tilde{\rho}_0\Big(\mathbf{k}-
\dfrac{\mathbf{q}}{2}\Big)\Big].
$$ \medskip

Hence, expression for current density has the following form

$$
\mathbf{j}(\mathbf{q},\omega,\bar\nu)=-\dfrac{e^2}{mc}\mathbf{A}(\mathbf{q})
\int \dfrac{d\mathbf{k}}{8\pi^3}\tilde{\rho}_0\Big(\mathbf{k}+
\dfrac{\mathbf{q}}{2}\Big)-\dfrac{e^2}{mc}\mathbf{A}(\mathbf{q})
\int \dfrac{d\mathbf{k}}{8\pi^3}\tilde{\rho}_0\Big(\mathbf{k}-
\dfrac{\mathbf{q}}{2}\Big)+
$$\medskip
$$
+e\hbar\int\dfrac{d\mathbf{k}}{4\pi^3m}
\left\langle\mathbf{k}+\dfrac{\mathbf{q}}{2}
\Big|\rho_1\Big|\mathbf{k}-\dfrac{\mathbf{q}}{2}\right\rangle.
$$ \medskip

First two members in this expression are equal each other
$$
\int \dfrac{d\mathbf{k}}{8\pi^3}\tilde{\rho}_0\Big(\mathbf{k}+
\dfrac{\mathbf{q}}{2}\Big)=
\int \dfrac{d\mathbf{k}}{8\pi^3}\tilde{\rho}_0\Big(\mathbf{k}-
\dfrac{\mathbf{q}}{2}\Big)=\dfrac{N}{2},
$$
where $N$ is the number density (concentration) of plasma.

Hence, the current density is equal
$$
{\bf j}({\bf q},\omega,\bar\nu)=-\frac{e^2N}{mc}{\bf A}({\bf q})+
e\hbar\int \dfrac{d{\bf k}}{4\pi^3m}{\bf k}
\left\langle\mathbf{k}+\dfrac{\mathbf{q}}{2}\Big|\rho_1\Big|{\bf k}-
\frac{\mathbf{q}}{2}\right\rangle.
\eqno{(2.2)}
$$

The first composed in (2.2) is not that other, as calibration current density.

By means of obvious replacement of variables in integral from (2.2)
expression (2.2) we can transform to the form
$$
{\bf j}({\bf q},\omega,\bar\nu)=-\frac{e^2N}{mc}{\bf A}({\bf q})+
e\hbar\int \dfrac{d{\bf k}}{4\pi^3m}{\bf k}
\left\langle\mathbf{k}\Big|\rho_1\Big|{\bf k}-{\bf q}\right\rangle.
\eqno{(2.3)}
$$

In the relation (2.3) subintegral expression is given by equality (1.12).
Substituting (1.12) in (2.3), we receive the following expression
for current density
$$
{\bf j}({\bf q},\omega,\bar\nu)=-\frac{e^2N}{mc}{\bf A}({\bf q})-
$$\medskip
$$-
\dfrac{e^2\hbar^2}{4\pi^3m^2c}
\int \mathbf{k}[\mathbf{k}\mathbf{A(q)}]\;\Xi(\mathbf{k,k-q})
(f_{{\bf k}}-f_{{\bf k-q}})d\mathbf{k}.
\eqno{(2.4)}
$$

Let's direct an axis $x$ along a vector ${\bf q} $, and an axis $y $
we direct lengthways vector $ {\bf A} $.
Then the previous vector expression (2.4) can be
rewrite in the form of three scalar
$$
{ j}_y({\bf q},\omega,\bar\nu)=-\frac{e^2N}{mc}{ A}({\bf q})-
\dfrac{e^2\hbar^2 A({\bf q})}{4\pi^3m^2c}\int { k}_y^2
\;\Xi(\mathbf{k,k-q})(f_{{\bf k}}-f_{{\bf k-q}}) d{\bf k}
$$
and
$$
{ j}_x({\bf q},\omega,\bar\nu)={ j}_z({\bf q},\omega,\bar\nu)=0.
$$

Obviously, that
$$
\int { k}_y^2\;\Xi(\mathbf{k,k-q})(f_{{\bf k}}-f_{{\bf k-q}})d{\bf k}
=\int { k}_z^2\;\Xi(\mathbf{k,k-q})(f_{{\bf k}}-f_{{\bf k-q}})d{\bf k}.
$$

Hence
$$
\int { k}_y^2\;\Xi(\mathbf{k,k-q})
(f_{{\bf k}}-f_{{\bf k-q}})d{\bf k}=
$$
$$=
\dfrac{1}{2}\int ({ k}_y^2+{
k}_z^2)\;\Xi(\mathbf{k,k-q})(f_{{\bf k}}-f_{{\bf k-q}})d{\bf k}=
$$
$$
=\dfrac{1}{2}\int ({\bf k}^2-{k}_x^2)\;\Xi(\mathbf{k,k-q})
(f_{{\bf k}}-f_{{\bf k-q}})d{\bf k}.
$$

From here we conclude, that expression for current density  we can present
in the following invariant form
$$
{\bf j}({\bf q},\omega,\bar\nu)=-\frac{Ne^2}{mc}{\bf A}({\bf q})-\hspace{6cm}
$$
$$\hspace{2cm}
-\dfrac{e^2\hbar^2}{8\pi^3m^2c}{\bf A}({\bf q})\int\;
\Xi(\mathbf{k,k-q})(f_{{\bf k}}-f_{{\bf k-q}}){\bf k}_\perp^2 d{\bf k},
\eqno{(2.5)}
$$
where
$$
{\bf k}_\perp^2={\bf k}^2-\Big(\dfrac{{\bf k}{\bf q}}{q}\Big)^2.
$$

Considering kernel decomposition $ \Xi(\mathbf {k, k-q}) $ on fraction
$$
\Xi(\mathbf{k,k-q})=\dfrac{1}{\E_{{\bf k}}-\E_{{\bf k-q}}}+
\dfrac{\hbar \omega}
{ \E_{{\bf k}}-\E_{{\bf k-q}}-\hbar[ \omega+i\bar\nu({\bf k}_1,{\bf k}_2)]},
$$
we present equality (2.5) in the following form
$$
{\bf j}({\bf q},\omega,\bar\nu)=-\frac{Ne^2}{mc}{\bf A}({\bf q})-
\dfrac{e^2\hbar^2}{8\pi^3m^2c}{\bf A}({\bf q})\int
\dfrac{f_{\bf k}-f_{\bf k-q}}{\E_{{\bf k}}-\E_{{\bf k-q}}}
{\bf k}_\perp^2 d{\bf k}-
$$
$$
-\dfrac{e^2\hbar^3\omega}{8\pi^3m^2c}{\bf A}({\bf q})\int
\dfrac{(f_{\bf k}-f_{\bf k-q}){\bf k}_\perp^2 d{\bf k}}
{(\E_{{\bf k}}-\E_{{\bf k-q}})
\{ \E_{{\bf k}}-\E_{{\bf k-q}}-\hbar[ \omega+i\bar\nu({\bf k}_1,{\bf k}_2)]\}}.
\eqno{(2.6)}
$$\medskip

First two members in the previous equality (2.6) do not depend on frequency
$ \omega $ also are defined by the dissipativity properties of a material
defined by frequency of collisions $ \nu ({\bf k}) $.
These members are universal para\-met\-res,
defining {\bf Landau diamagnetism}.

\begin{center}
\bf 3. Transversal electric conductivity and dielectric permeability
\end{center}

Considering communication of vector potential with intensity of an
electromagnetic field, and also communication of density of a current
with electric field, on the basis of the previous
equality (2.5) we receive the following expression of an invariant form
for the transversal electric con\-duc\-ti\-vity
$$
\sigma_{tr}(\mathbf{q},\omega,\bar\nu)=
\dfrac{ie^2N}{m\omega}\Bigg[1+\dfrac{\hbar^2}{8\pi^3mN}
\int \Xi(\mathbf{k,k-q})(f_{\bf k}-f_{\bf k-q}){\bf k}_\perp^2 d\mathbf{k}
\Bigg].
\eqno{(3.1)}
$$

Let's take advantage of definition of transversal dielectric permeability
$$
\varepsilon_{tr}(\mathbf{q},\omega,\nu)=
1+\dfrac{4\pi i}{\omega}\sigma_{tr}(\mathbf{q},\omega,\nu).
\eqno{(3.2)}
$$

Taking into account (3.1) and equality (3.2) we will write expression
for the transversal dielectric permeability
$$
\varepsilon_{tr}(\mathbf{q},\omega,\bar\nu)=1-\dfrac{\omega_p^2}{\omega^2}
\Big[1+\dfrac{\hbar^2}{8\pi^3mN}\int \Xi(\mathbf{k,q})(f_{\mathbf{k}}-
f_{\mathbf{k-q}})\mathbf{k}_\perp^2 d\mathbf{k}\Big].
\eqno{(3.3)}
$$

Here $\omega_p$ is the plasma (Langmuir) frequency,
$\omega_p^2=4\pi e^2N/m$.

From equality (3.3) it is visible, that one of equalities named a rule
$f$-sums is carried out (see, for example, \cite{Dressel}, \cite{Pains} and
\cite{Martin}) for the transversal dielectric permeability. This rule is
expressed by the formula (4.200) from the monography \cite{Pains}
$$\boxed{
\int\limits_{-\infty}^{\infty}\varepsilon_{tr}(\mathbf{q},\omega,\nu)\omega
d\omega=\pi \omega_p^2}.
$$

Let's notice, that the kernel from subintegral expression from (3.1) can be
we can present in the form of decomposition on partial fractions
$$
\dfrac{\E_{\mathbf{k}}-\E_{\mathbf{k-q}}-i\hbar \bar\nu({\bf k},{\bf k-q})}
{(\E_{\mathbf{k}}-\E_{\mathbf{k-q}})\{\E_{\mathbf{k}}-\E_{\mathbf{k-q}}-\hbar
[\omega+i\bar\nu({\bf k},{\bf k-q})]\}}=
$$
$$
=\dfrac{i \bar\nu({\bf k},{\bf k-q})}{\omega+i\bar\nu({\bf k},{\bf k-q})}\cdot
\dfrac{1}
{\E_{\mathbf{k}}-\E_{\mathbf{k-q}}}+
$$
$$+
\dfrac{\omega}{\omega+i\bar\nu({\bf k},{\bf k-q})}\cdot\dfrac{1}
{\E_{\mathbf{k}}-\E_{\mathbf{k-q}}-\hbar [\omega+i\bar\nu({\bf k},{\bf k-q})]}.
$$

Hence, for transversal electric conductivity and dielectric
permeability we have explicit representations
$$
\sigma_{tr}(\mathbf{q},\omega,\bar\nu)=
\dfrac{ie^2N}{m\omega}\Bigg[1+\dfrac{\hbar^2}{8\pi^3mN}
\int \dfrac{i \bar\nu({\bf k},{\bf k-q})}{\omega+i\bar\nu({\bf k},{\bf k-q})}
\cdot\dfrac{f_{\bf k}-f_{\bf k-q}}{\E_{{\bf k}}-\E_{{\bf k-q}}}{\bf k}_\perp^2
d{\bf k}+
$$
$$
+\dfrac{\hbar^2\omega}{8\pi^3mN}\int
\dfrac{1}{\omega+i\bar\nu({\bf k},{\bf k-q})}\cdot
\dfrac{(f_{\bf k}-f_{\bf k-q}){\bf k}_\perp^2d{\bf k}}{\E_{{\bf k}}-\E_{{\bf k-q}}
-\hbar [\omega+i\bar\nu({\bf k},{\bf k-q})]}\Bigg],
\eqno{(3.2)}
$$\medskip
and
$$
\varepsilon_{tr}(\mathbf{q},\omega,\bar\nu)=1-\dfrac{\omega_p^2}{\omega^2}
\Bigg[1+\dfrac{\hbar^2}{8\pi^3mN}
\int \dfrac{i \bar\nu({\bf k},{\bf k-q})}{\omega+i\bar\nu({\bf k},{\bf k-q})}
\cdot\dfrac{f_{\bf k}-f_{\bf k-q}}{\E_{{\bf k}}-\E_{{\bf k-q}}}{\bf k}_\perp^2
d{\bf k}+
$$
$$
+\dfrac{\hbar^2\omega}{8\pi^3mN}\int
\dfrac{1}{\omega+i\bar\nu({\bf k},{\bf k-q})}\cdot
\dfrac{(f_{\bf k}-f_{\bf k-q}){\bf k}_\perp^2d{\bf k}}{\E_{{\bf k}}-\E_{{\bf k-q}}
-\hbar [\omega+i\bar\nu({\bf k},{\bf k-q})]}\Bigg].
\eqno{(3.3)}
$$\medskip

If to enter designations
$$
J_{\bar\nu}=\dfrac{\hbar^2}{8\pi^3mN}
\int \dfrac{i \bar\nu({\bf k},{\bf k-q})}{\omega+i\bar\nu({\bf k},{\bf k-q})}
\cdot\dfrac{f_{\bf k}-f_{\bf k-q}}{\E_{{\bf k}}-\E_{{\bf k-q}}}{\bf k}_\perp^2
d{\bf k}
$$
and
$$
J_\omega=\dfrac{\hbar^2\omega}{8\pi^3mN}\int
\dfrac{1}{\omega+i\bar\nu({\bf k},{\bf k-q})}\cdot
\dfrac{(f_{\bf k}-f_{\bf k-q}){\bf k}_\perp^2d{\bf k}}{\E_{{\bf k}}-\E_{{\bf k-q}}
-\hbar [\omega+i\bar\nu({\bf k},{\bf k-q})]},
$$
then expression (3.2) for electric conductivity and (3.3) for
dielectric perme\-abi\-lity will be transformed to the following form
$$
\sigma_{tr}(\mathbf{q},\omega,\bar\nu)=\dfrac{ie^2N}{m\omega}\big(1+
J_{\bar\nu}+J_\omega \big)
\eqno{(3.4)}
$$
and
$$
\varepsilon_{tr}(\mathbf{q},\omega,\bar\nu)=1-\dfrac{\omega_p^2}{\omega^2}
\big(1+J_{\bar\nu}+J_\omega \big).
\eqno{(3.5)}
$$

Let's notice, that in case of constant frequency of collisions we have
$\bar\nu({\bf k},{\bf k-q})=\nu$ and formulas (3.4) and (3.5)
will be transformed in corresponding formulas from our work \cite{Trans2}
$$\boxed{
\dfrac{\sigma_{tr}(\mathbf{q},\omega,\nu)}{\sigma_0}=
\dfrac{i \nu}{\omega}\Big(1+\dfrac{\omega I_\omega+i \nu I_\nu}{\omega+i \nu}
\Big)}
$$
and
$$\boxed{
\varepsilon_{tr}(\mathbf{q},\omega,\nu)=
1-\dfrac{\omega_p^2}{\omega^2}
\Big(1+\dfrac{\omega I_\omega+i \nu I_\nu}{\omega+i \nu}
\Big)}.
$$

Here
$$
I_\nu=\dfrac{\hbar^2}{8\pi^3mN}
\int \dfrac{f_{{\bf k}}-f_{{\bf k-q}}}{\E_{{\bf k}}-\E_{{\bf k-q}}}
\mathbf{k}^2_\perp d{\bf k}
$$
and
$$
I_\omega=\dfrac{\hbar^2}{8\pi^3mN}\int
\dfrac{(f_{{\bf k}}-f_{{\bf k-q}})\mathbf{k}^2_\perp d{\bf k}}
{\E_{{\bf k}}-\E_{{\bf k-q}}-\hbar(\omega+i\nu)}.
$$

Let now $\nu({\bf k})\equiv 0$, i.e. $\bar\nu({\bf k,k-q})=
\bar \nu({\bf k+q},{\bf k}) \equiv 0$. In this case
dielectric permeability equals
$$
\varepsilon_{tr}({\bf q},\omega)=1-$$$$-\dfrac{\omega_p^2}{\omega^2}
\Big[1+\dfrac{\hbar^2}{8\pi^3mN}\int \Big(\dfrac{1}{\E_{{\bf k}}-
\E_{{\bf k-q}}-\hbar \omega}- \dfrac{1}{\E_{{\bf k+q}}-
\E_{{\bf k}}-\hbar \omega}\Big)f_{{\bf k}}{\bf k}_\perp^2d{\bf k}\Big].
$$ \bigskip

Let's return to the case of variable frequency of collisions.
Integrals $J_{\bar\nu}$ and  $J_\omega$ we can transform to the following
form
$$
J_{\bar\nu}=\dfrac{i\hbar^2}{8\pi^3mN}
\int\Bigg[\dfrac{\bar\nu({\bf k},{\bf k-q})}{[\omega+i\bar\nu({\bf k},{\bf k-q})]
\{\E_{{\bf k}}-\E_{{\bf k-q}}\}}-
$$
$$
-\dfrac{\bar\nu({\bf k+q},{\bf k})}{[\omega+i\bar\nu({\bf k+q},{\bf k})]
\{\E_{{\bf k+q}}-\E_{{\bf k}}\}}
\Bigg]f_{{\bf k}}{\bf k}_\perp^2d{\bf k}
\eqno{(3.8)}
$$
and
$$
J_{\omega}=\dfrac{\omega\hbar^2}{8\pi^3mN}
\int\Bigg[\dfrac{1}{[\omega+i\bar\nu({\bf k},{\bf k-q})]
\{\E_{{\bf k}}-\E_{{\bf k-q}}-\hbar [\omega+i\bar\nu({\bf k},{\bf k-q})]\}}-
$$
$$
-\dfrac{1}{[\omega+i\bar\nu({\bf k+q},{\bf k})]
\{\E_{{\bf k+q}}-\E_{{\bf k}}-\hbar [\omega+i\bar\nu({\bf k+q},{\bf k})]\}}
\Bigg]f_{{\bf k}}{\bf k}_\perp^2d{\bf k}
\eqno{(3.9)}
$$

\begin{center}
\bf 4. Degenerate plasma
\end{center}

Instead of the vector $\mathbf{k}$ we will enter the dimensionless
vector $\mathbf {K}$ by following equality
$\mathbf{K}=\dfrac{\mathbf{k}}{k_F}$, \quad
$k_F=\dfrac{p_F}{\hbar}$, where $k_F$ is the Fermi wave number,
$p_F=mv_F$ is the electron momentum on the Fermi surface, $v_F$ is the
electron velocity on the Fermi surface.

Then
$$
(k^2-k_x^2)d^3k=k_F^5(K^2-K_x^2)d^3K=k_F^5K_\perp^2d^3K,
$$
where
$$
K_\perp^2=K^2-K_x^2=K_y^2+K_z^2.
$$

Further we will consider the case of degenerate plasmas.
Then we have
$$
\Big(\dfrac{mv_F}{\hbar}\Big)^3 \equiv \Big(\dfrac{p_F}{\hbar}\Big)^3\equiv
k_F^3=3\pi^2N.
$$

Hence,
$$
(k^2-k_x^2)d^3k=3\dfrac{\pi^2 N m^2v_F^2}{\hbar^{2}}(K^2-K_x^2)d^3K.
$$

Absolute  Fermi---Dirac's distribution  $f_{\mathbf{k}}$
for degenerate plasma transforms into Fermi distribution
$f_{\mathbf{k}}=\Theta(\mathbf{k})\equiv\Theta(\E_F-\E_{\mathbf{k}})$.
Here $\Theta(x)$ is the Heaviside step function,
$$
\Theta(x)=\left\{\begin{array}{c}
                   1,\qquad x>0, \\
                   0,\qquad x<0.
                 \end{array}\right.
$$

Here $\E_F=\dfrac{p_F^2}{2m}$ is the electron energy on Fermi surface.

Calculation of transversal electric conductivity can be spent on any of
formulas (5.4) - (5.6). Let's calculate the integrals entering into
expression (5.6). The first integral is special case of the second.
Therefore at first we will calculate the second integral.
Energy $\E_{\mathbf {k}} $ we will express through Fermi's energy. We have
$$
\E_{\mathbf{k}}=\dfrac{\hbar^2\mathbf{k}^2}{2m}=\dfrac{\hbar^2k_F^2}{2m}
\mathbf{K}^2=\dfrac{p_F^2}{2m}\mathbf{K}^2=
\E_F\mathbf{K}^2\equiv \E_{\mathbf{K}}.
$$

In the same way we receive
$$
\E_{\mathbf{k-q}}=\dfrac{\hbar^2(k_F\mathbf{K}-\mathbf{q})^2}{2m}=
\dfrac{\hbar^2k_F^2}{2m}\Big(\mathbf{K}-\dfrac{\mathbf{q}}{k_F}\Big)^2.
$$

Further we introduce dimensionless wave vector
$\mathbf{Q}=\dfrac{\mathbf{q}}{k_F}$. Then
$$
\E_{\mathbf{k-q}}=\dfrac{\hbar^2k_F^2}{2m}\Big(\mathbf{K}-\mathbf{Q}\Big)^2=
\E_F(\mathbf{K-Q})^2=\E_{{\bf K-Q}}.
$$

We notice that
$$
\E_{\mathbf{K}}-\E_{\mathbf{K-Q}}=\E_F\mathbf{K}^2-\E_F(\mathbf{K-Q})^2=
\E_F[2K_xQ-Q^2]=
$$
$$
=2Q\E_F(K_x-\dfrac{Q}{2}).
$$

Besides
$$
\E_{\mathbf{K}}-\E_{\mathbf{K-Q}}-\hbar [\omega+i\bar\nu({\bf K},{\bf K-Q})]=
2\E_FQ\Big(K_x-\dfrac{z^-}{Q}-\dfrac{Q}{2}),
$$
$$
\E_{\mathbf{K+Q}}-\E_{\mathbf{K}}-\hbar [\omega+i\bar\nu({\bf K+Q},{\bf K})]=
2\E_FQ\Big(K_x-\dfrac{z^+}{Q}+\dfrac{Q}{2}),
$$
where
$$
z^{\pm}=x+iy^{\pm},\qquad x=\dfrac{\omega}{k_Fv_F},
$$
$$
y^-=\dfrac{\bar\nu({\bf K},{\bf K-Q})}{k_Fv_F},\qquad
y^+=\dfrac{\bar\nu({\bf K+Q},{\bf K})}{k_Fv_F}.
$$

Let's consider the special case, when frequency of collisions
is proportional to the module of a wave vector
$$
\nu({\bf k})=\nu_0|{\bf k}|.
$$

Then
$$
\bar\nu({\bf k,k-q})=\dfrac{\nu({\bf k})+\nu({\bf k-q})}{2}=
\dfrac{\nu_0}{2}\Big(|{\bf k}|+|{\bf k-q}|\Big),
$$
and
$$
\bar\nu({\bf k+q,k})=\dfrac{\nu({\bf k+q})+\nu({\bf k})}{2}=
\dfrac{\nu_0}{2}\Big(|{\bf k+q}|+|{\bf k}|\Big).
$$

The quantity $ \nu_0$ we take in the form $\nu_0=\dfrac{\nu}{k_F}$, where
$k_F$ is the Fermi wave number, $k_F=\dfrac{mv_F}{\hbar}$,
$ \hbar $ is the Planck's constant, $v_F$ is the Fermi electron velocity. Now
we have
$$
\nu({\bf k})=\dfrac{\nu}{k_F}|{\bf k}|.
\eqno{(4.1)}
$$

Let's notice, that on Fermi's surface, i.e. at $k=k_F $:
$ \nu (k_F) = \nu $. So, further in previous formulas frequency
collisions according to (4.1) it is equal
$$
\bar\nu({\bf k,k-q})=\dfrac{\nu}{2k_F}
\big(|{\bf k}|+|{\bf k-q}|\big)=\dfrac{\nu}{2}\Big(|{\bf K}|+|{\bf K-Q}|\Big)=
\bar\nu({\bf K,K-Q}),
$$
$$
\bar\nu({\bf k+q,k})=\dfrac{\nu}{2k_F}
\big(|{\bf k+q}|+|{\bf k}|\big)=\dfrac{\nu}{2}\Big(|{\bf K+Q}|+|{\bf K}|\Big)=
\bar\nu({\bf K+Q,K}).
$$

Hence, quantities $z^{\pm}$ are equal
$$
z^{\pm}=x+iy\rho^{\pm}, \qquad y=\dfrac{\nu}{k_Fv_F},
$$
$$
\rho^{\pm}=\dfrac{1}{2}\Big(\sqrt{K_x^2+K_y^2+K_z^2}+
\sqrt{(K_x\pm Q)^2+K_y^2+K_z^2}\Big).
$$

Now integrals $J_{\bar\nu} $ and $J_\omega $ are accordingly equal
$$
J_{\bar\nu}=\dfrac{3iy}{8\pi Q}\int\Bigg(\dfrac{\rho^-}{(x+iy\rho^-)
(K_x-Q/2)}-\hspace{5cm}$$$$\hspace{3cm}-\dfrac{\rho^+}{(x+iy\rho^+)
(K_x+Q/2)}\Bigg)f_{{\bf K}}{\bf K}_\perp^2d^3K,
\eqno{(4.2)}
$$
and
$$
J_{\omega}=\dfrac{3x}{8\pi Q}\int\Bigg(\dfrac{1}{(x+iy\rho^-)
(K_x-z^-/Q-Q/2)}-\hspace{4cm}$$$$\hspace{4cm}-\dfrac{1}{(x+iy\rho^+)
(K_x-z^+/Q+Q/2)}\Bigg)f_{{\bf K}}{\bf K}_\perp^2d^3K.
\eqno{(4.3)}
$$

Three-dimensional integrals (4.2) and (4.3) after passing to polar
coor\-di\-na\-tes in a plane $(K_y, K_z) $ are easily reduced to the double
$$
J_{\bar\nu}=\dfrac{3iy}{4Q}\int\limits_{-1}^{+1}dK_x
\int\limits_{0}^{\sqrt{1-K_x^2}}\Bigg(\dfrac{\rho^-}{(x+iy\rho^-)
(K_x-Q/2)}-\hspace{5cm}
$$
$$
\hspace{3cm}-\dfrac{\rho^+}{(x+iy\rho^+)
(K_x+Q/2)}\Bigg)r^3dr,
\eqno{(4.4)}
$$
and
$$
J_{\omega}=\dfrac{3x}{4 Q}\int\limits_{-1}^{+1}dK_x
\int\limits_{0}^{\sqrt{1-K_x^2}}\Bigg(\dfrac{1}{(x+iy\rho^-)
(K_x-z^-/Q-Q/2)}-\hspace{4cm}
$$
$$
\hspace{4cm}-\dfrac{1}{(x+iy\rho^+)
(K_x-z^+/Q+Q/2)}\Bigg)r^3dr.
\eqno{(4.5)}
$$

Let's notice, that in case of constant frequency of collisions
$\rho^{\pm} =1$ and formulas (4.4) and (4.5) pass in the following
$$
J_\omega=\dfrac{3x}{16(x+iy)}T(Q,z),\qquad
T(Q,z)=\int\limits_{-1}^{1}\dfrac{(1-K_x^2)^2dK_x}{(K_x-z/Q)-(Q/2)^2},
$$
$$
J_{\nu}=\dfrac{3iy}{16(x+iy)T(Q,0)}, \qquad z=x+iy.
$$

By means of these expressions we receive known formulas for
electric conductivity and dielectric permeability
of quantum collisional degenerate plasmas with constant
frequency of collisions of particles \cite{Trans2}
$$
\dfrac{\sigma_{tr}}{\sigma_0}=\dfrac{iy}{x}\Big[1+\dfrac{3}{16}
\dfrac{xT(Q,z)+iyT(Q,0)}{x+iy}\Big]
$$
and
$$
\varepsilon_{tr}=1-\dfrac{\omega_p^2}{\omega^2}\Big[1+\dfrac{3}{16}
\dfrac{xT(Q,z)+iyT(Q,0)}{x+iy}\Big].
$$

On Figs. 1 -- 8 we will present comparison real and imaginary
parts of dielectric function.

\begin{center}
\bf 5. Conclusion
\end{center}

In the present work formulas for the transversal electric
conductivity and dielectric permeability into quantum collisional
plasma are deduced.
Frequency of collisions of particles depends arbitrarily on a wave vector.
For this purpose the kinetic equation with integral of collisions
in form of relaxation model in momentum space is used.
The case of degenerate Fermi plasma is allocated and investigated.
The special case of frequency of collisions
proportional to the module of a wave vector is considered.
The graphic analysis of the real and imaginary parts  of
dielectric function is made.

\begin{figure}[t]\center
\includegraphics[width=16.0cm, height=10cm]{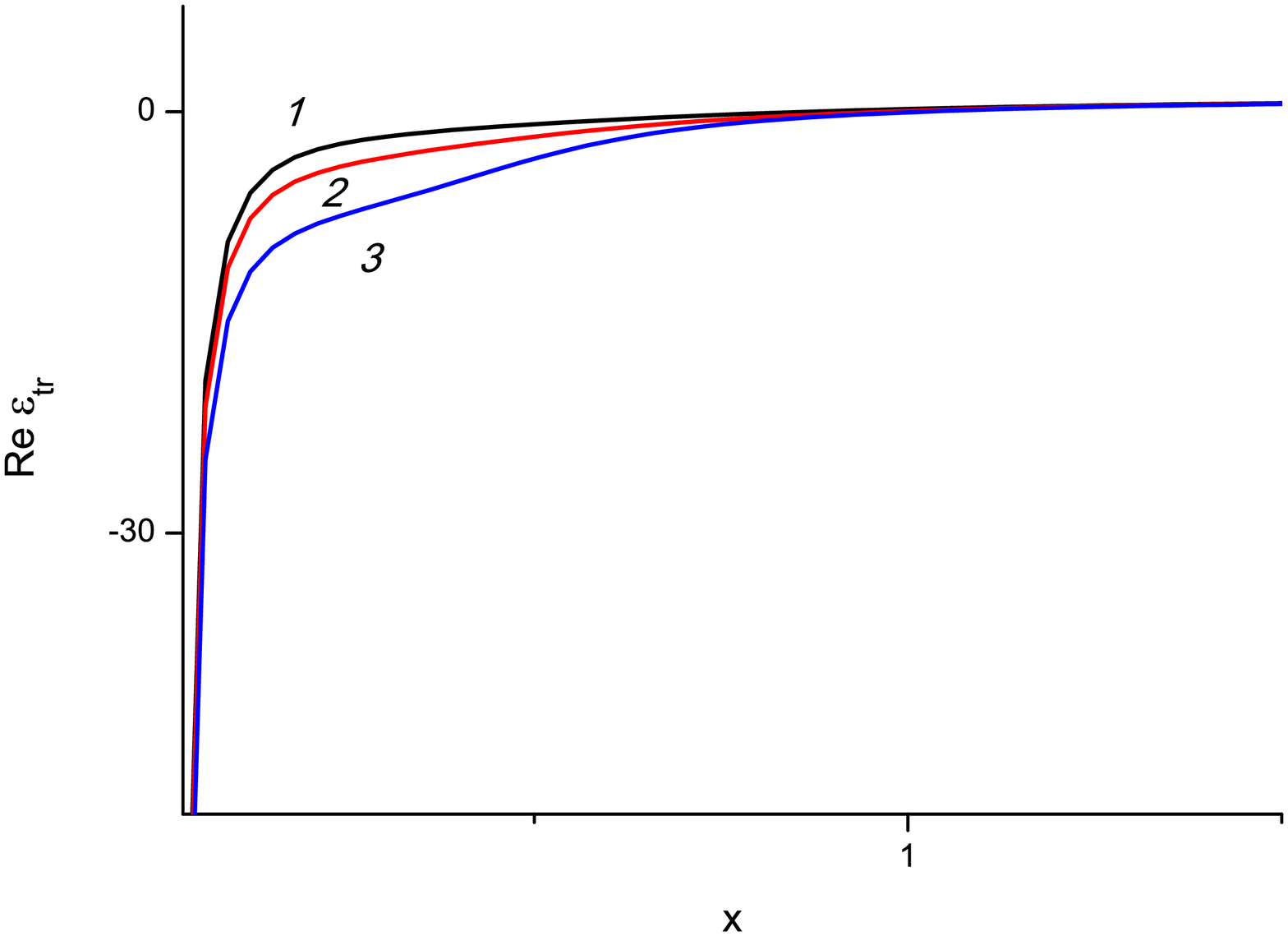}
\center{Fig. 1. Real part of dielectric function,
$x_p=1, Q=1$. Curves 1,2,3 correspond to values $y=0.5,0.3,0.1$.}
\includegraphics[width=17.0cm, height=10cm]{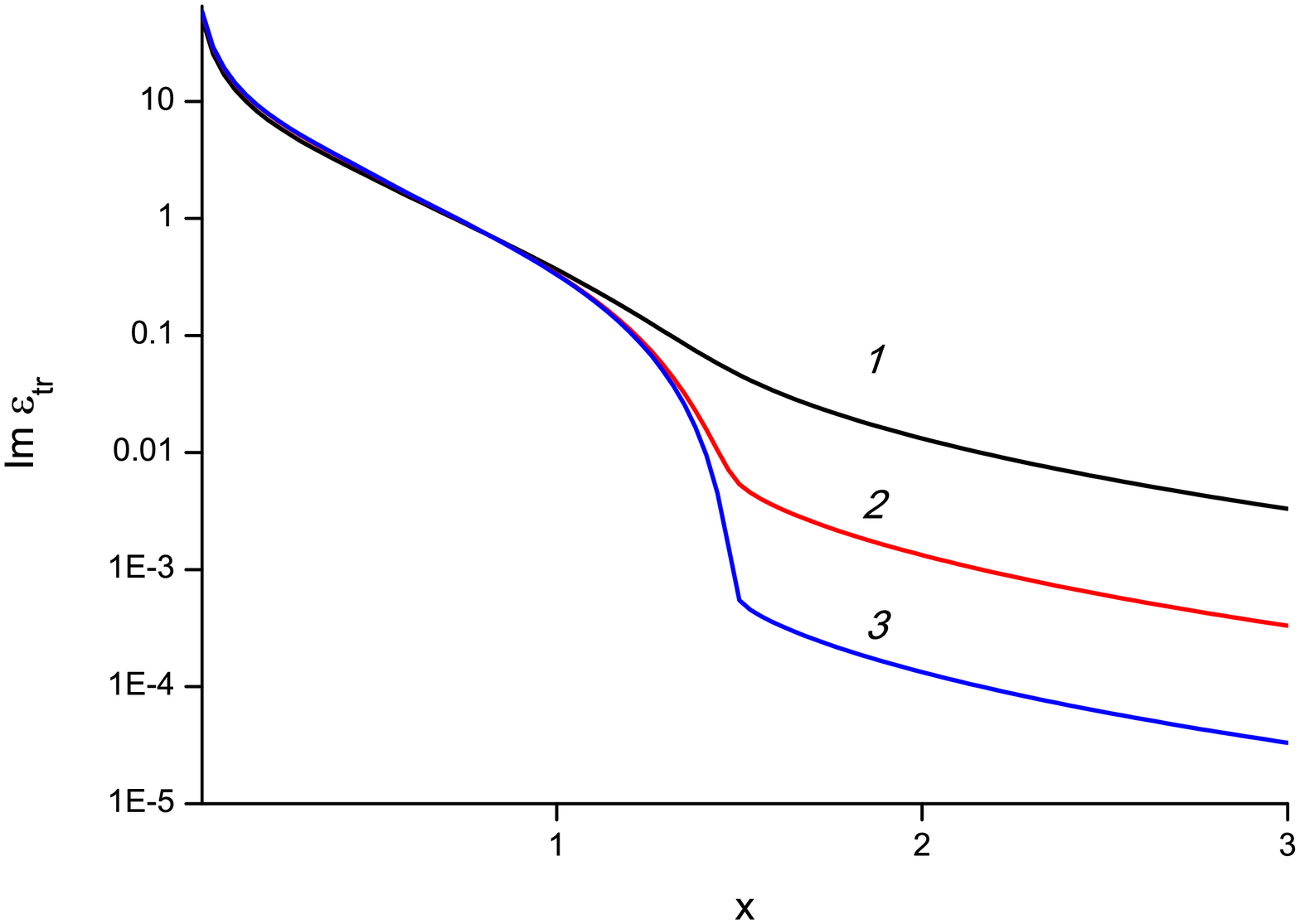}
\center{Fig. 2. Imaginare part of dielectric function,
$x_p=1, Q=1$. Curves 1,2,3 correspond to values $y=0.5,0.3,0.1$.}
\end{figure}

\begin{figure}[h]\center
\includegraphics[width=16.0cm, height=10cm]{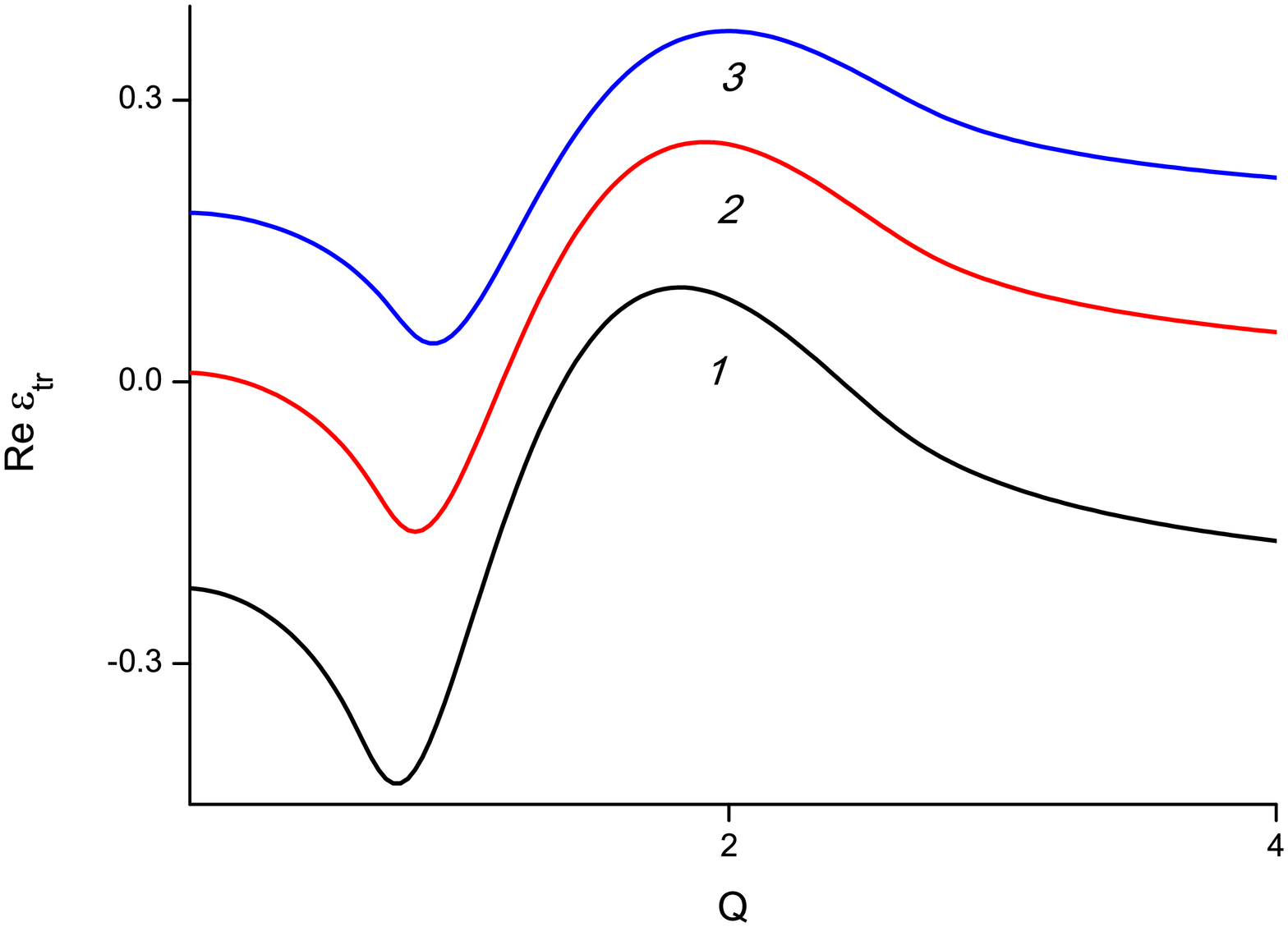}
\center{Fig. 3. Real part of dielectric function,
$x_p=1, y=0.1$. Curves 1,2,3 correspond to values $x=0.9,1.0,1.1$.}
\includegraphics[width=17.0cm, height=10cm]{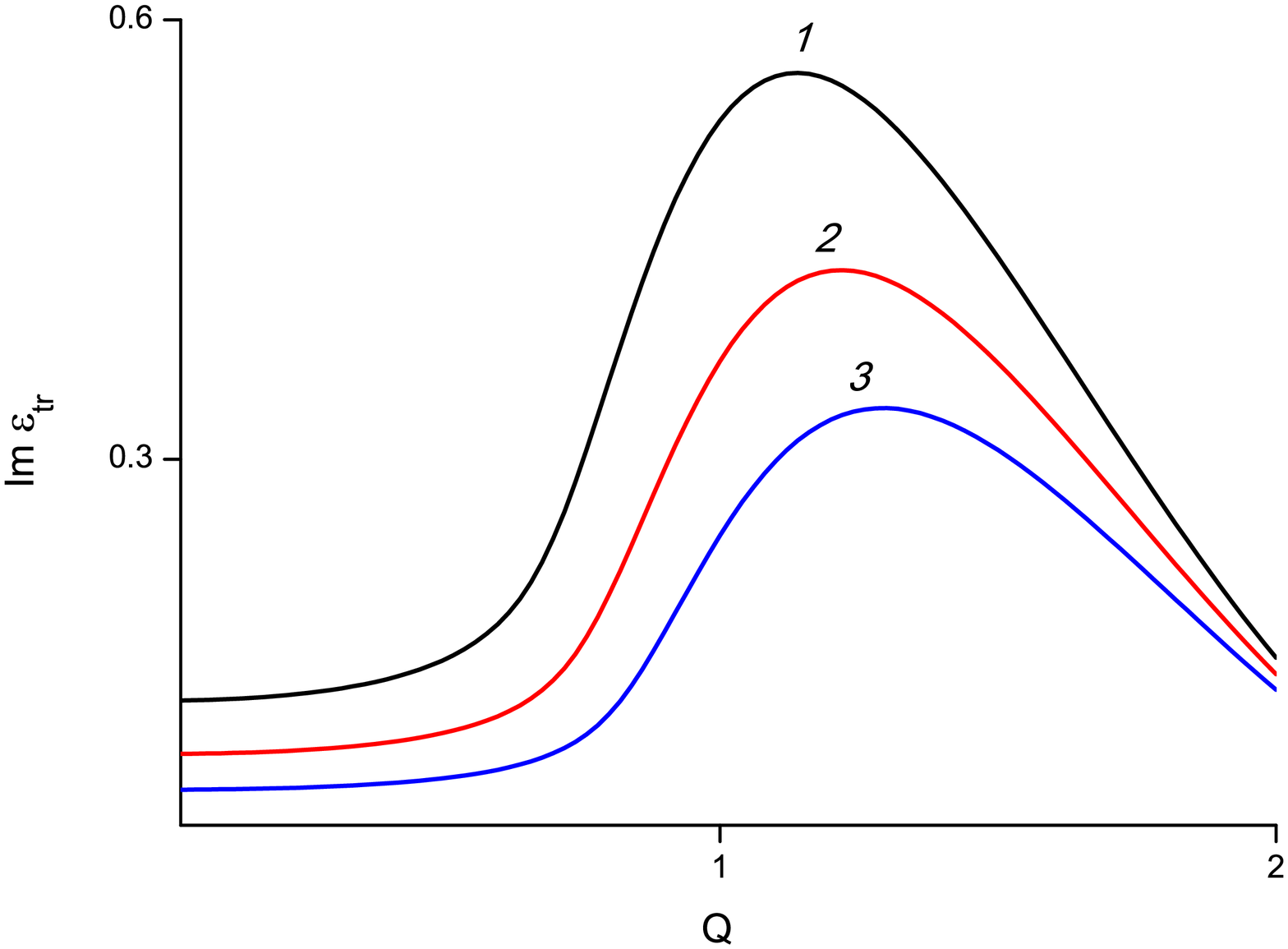}
\center{Fig. 4. Imaginare part of dielectric function,
$x_p=1, y=0.01$. Curves 1,2,3 correspond to values $x=0.9,1.0,1.1$.}
\end{figure}

\begin{figure}[h]\center
\includegraphics[width=16.0cm, height=10cm]{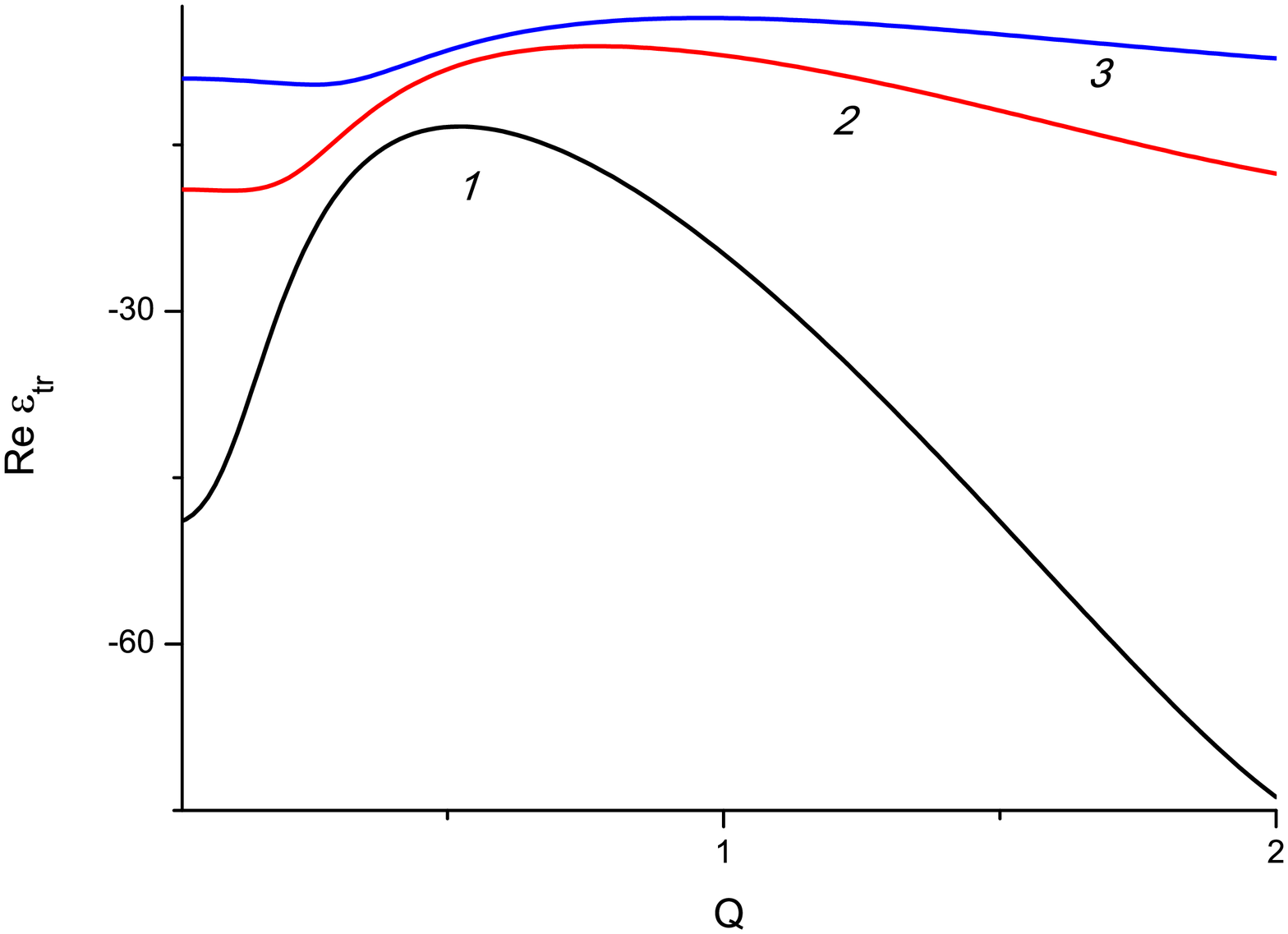}
\center{Fig. 5. Real part of dielectric function,
$x_p=1, y=0.1$. Curves 1,2,3 correspond to values $x=0.1,0.2,0.3$.}
\includegraphics[width=17.0cm, height=10cm]{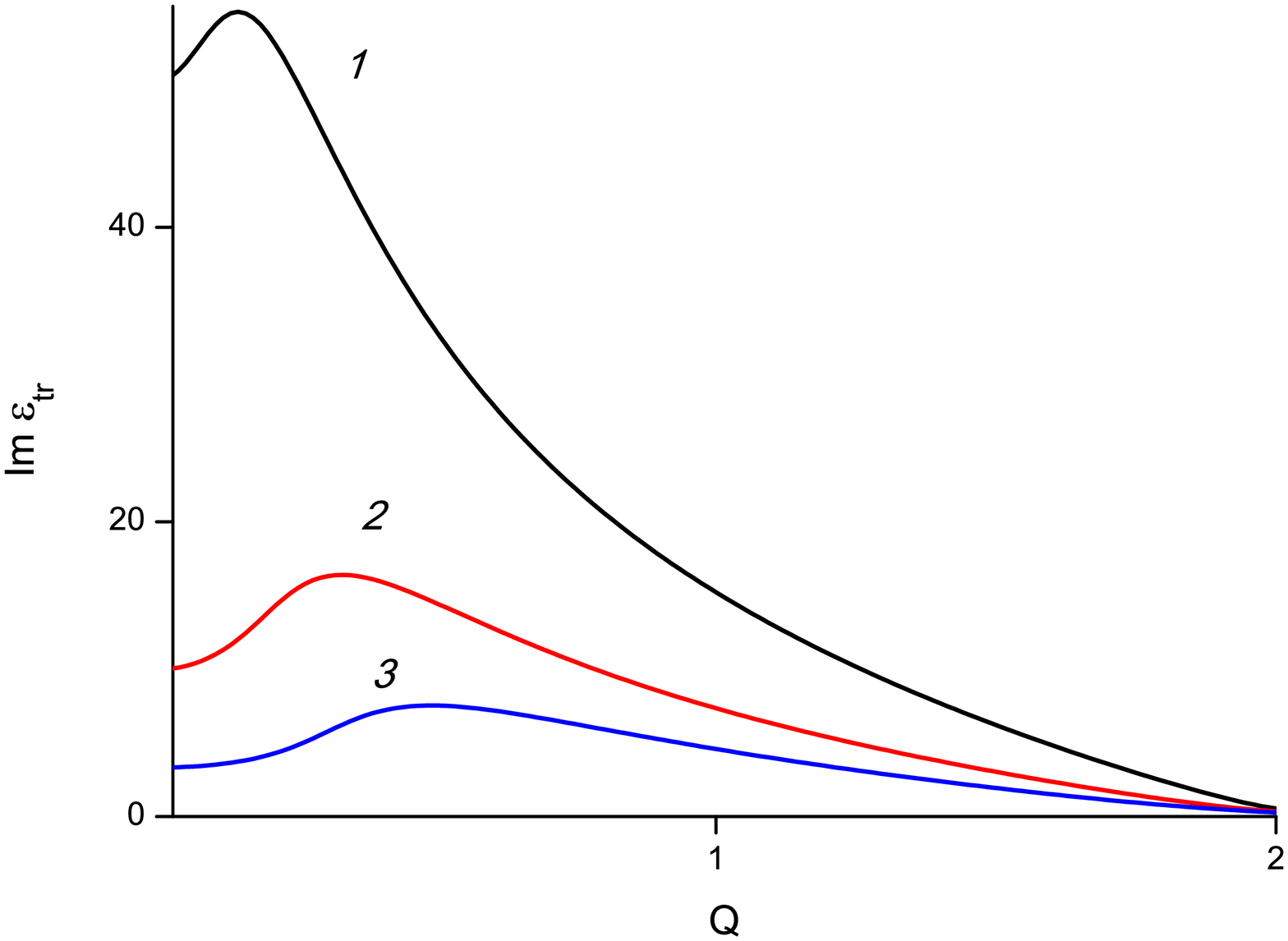}
\center{Fig. 6. Imaginare part of dielectric function,
$x_p=1, y=0.1$. Curves 1,2,3 correspond to values $x=x=0.1,0.2,0.3$.}
\end{figure}

\begin{figure}[h]\center
\includegraphics[width=16.0cm, height=10cm]{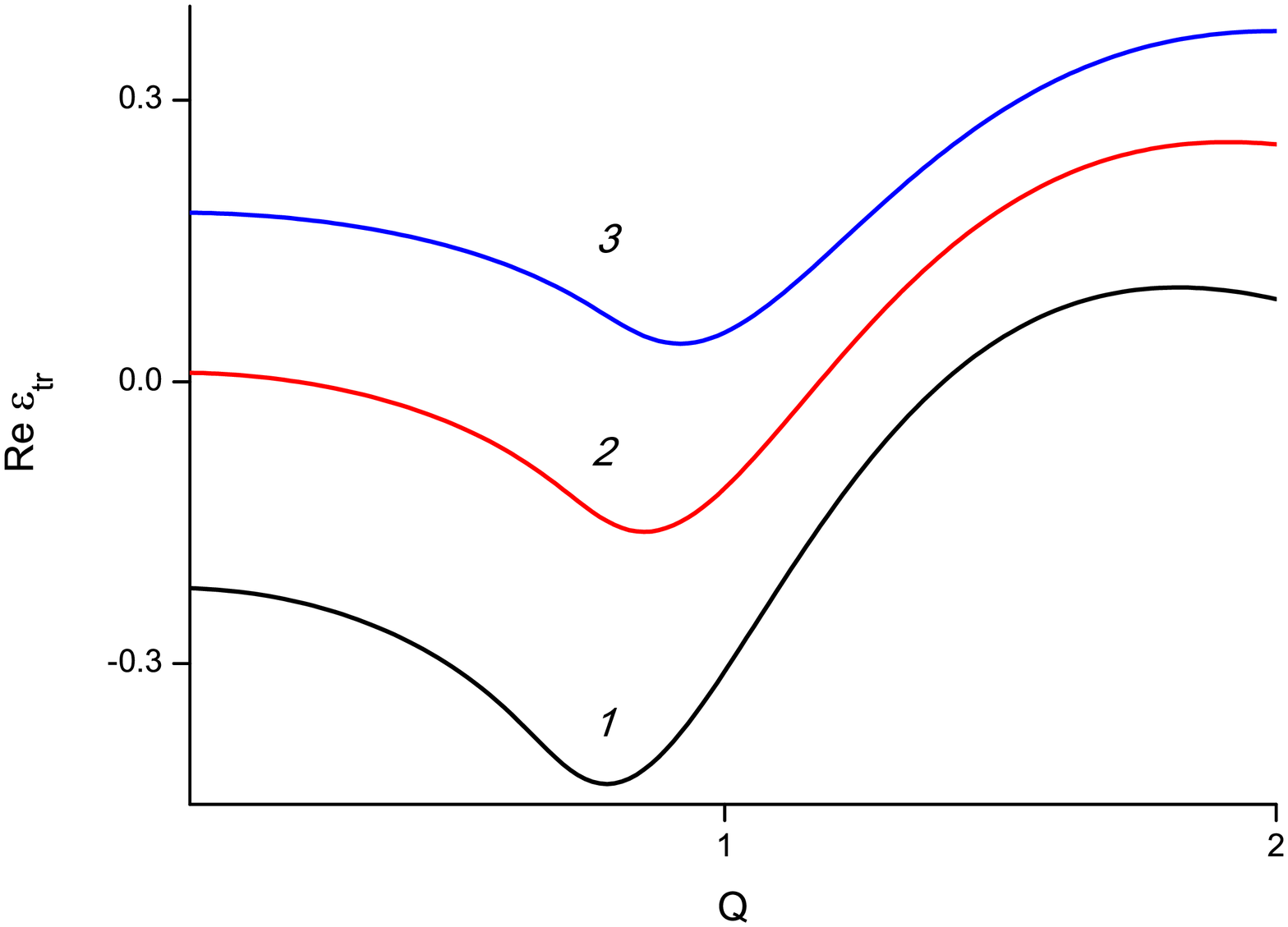}
\center{Fig. 7. Real part of dielectric function,
$x_p=1, y=0.1$. Curves 1,2,3 correspond to values $x=0.9,1.0,1.1$.}
\includegraphics[width=17.0cm, height=10cm]{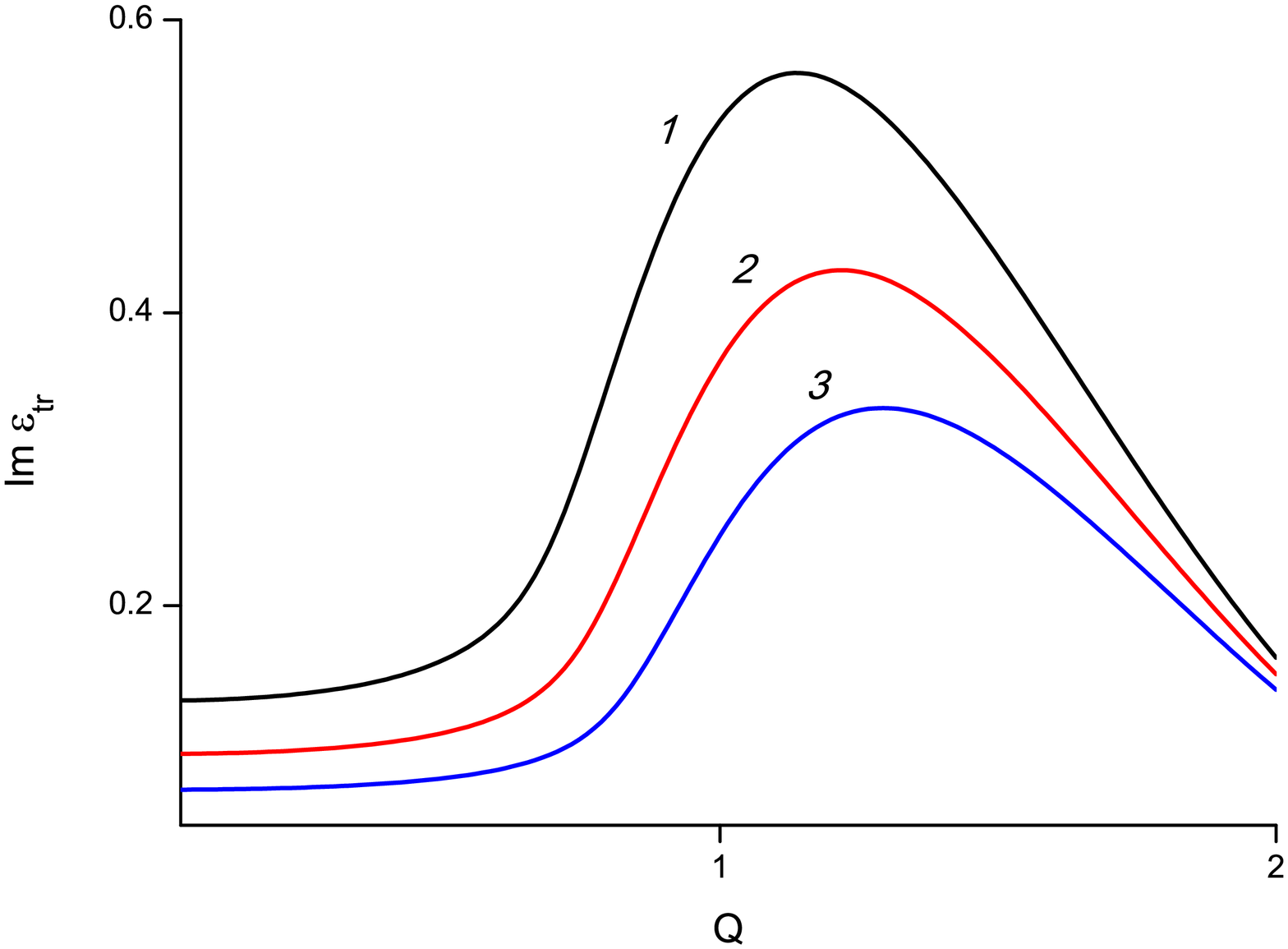}
\center{Fig. 8. Imaginare part of dielectric function,
$x_p=1, y=0.1$. Curves 1,2,3 correspond to values $x=0.9,1.0,1.1$.}
\end{figure}

\end{document}